Accepted at 2026 IEEE 17th Annual Ubiquitous Computing, Electronics & Mobile Communication Conference (UEMCON) - 2026 IEEE UEMCON

# Bridging the Gap in ECG-Based Emotion Recognition: A Unified Evaluation of Deep Learning Models

Timothy C Sweeney-Fanelli
*Clarkson University, USA*
*Affects AI LLC, USA*
tfanelli@clarkson.edu

Ajan Ahmed
*Clarkson University, USA*
aahmed@clarkson.edu

Masudul Imtiaz
*Clarkson University, USA*
mimtiaz@clarkson.edu

***Abstract*—Deep learning has led to numerous proposed architectures for Automated Emotion Recognition (AER) from electrocardiogram (ECG) data, but inconsistencies in pre-processing, training, and evaluation make direct comparisons difficult. Most studies train and validate models on individual datasets collected under homogeneous conditions, limiting variability and raising concerns about generalizability. Cross-dataset validation is sometimes used but primarily assesses model adaptability rather than true generalization. This study presents a comparative analysis of prominent deep learning architectures in AER, emphasizing model generalization over dataset adaptability. To enable this benchmark, we introduce two open-source frameworks: Affective Research on Representations and Classifications (ARRC), a standardized benchmarking toolkit, and Affective Research Dataset Toolkit (ARDT), a framework for inter-dataset training and validation. Using ARDT, we consolidate three publicly available AER datasets—CUADS, ASCERTAIN, and DREAMER—into a single dataset, increasing variability in sensor types, recording conditions, and participant demographics. We then use ARRC to evaluate three widely studied deep learning models and two CNN baselines through hyperparameter optimization and 10-fold cross-validation. Our findings provide insights into the trade-offs between classification accuracy and model complexity, establishing a reproducible benchmark for AER research. All source code for ARRC, ARDT, and model evaluation is publicly available to ensure transparency and facilitate further research.**

## I. Introduction

Deep Learning has emerged as a leading approach in Automated Emotion Recognition (AER) [1] from physiological signals, particularly electrocardiogram (ECG) data [2]. Numerous deep learning architectures have been proposed in recent AER literature, each reporting high classification accuracies [3]–[5]. However, comparing these models remains challenging [6]. Variations in dataset pre-processing, training protocols, evaluation metrics, and hyperparameter selection make direct comparisons unreliable [7]. Additionally, many published studies lack crucial implementation details—such as kernel and weight initializations and regularization strategies—which make it difficult for researchers to replicate results and verify model performance [8]. Many AER studies train and validate models on individual datasets collected under consistent conditions, limiting dataset variability and raising concerns about generalization and broader applicability of results [9]. While cross-dataset validation is often used, it assesses the model's ability to adapt to different datasets rather than its generalization across diverse real-world data.

This study provides a comparative benchmark of several prominent deep learning architectures proposed in recent AER literature, including the Temporal CNN auto-encoder [5], PETSFCNN [3], and CNN-LSTM [4]. As each model is based on a CNN architecture, we also include a simple 1-D CNN and a CNN-based residual network model for baseline comparison. By establishing a standardized evaluation framework, we are able to conduct a direct, controlled comparison of these deep learning models under identical signal pre-processing, hyperparameter tuning, and validation conditions. This approach ensures that performance differences between models stem from architectural variations rather than inconsistencies in data handling or training methodology.

We perform inter-dataset training and validation using three publicly available AER datasets, ASCERTAIN [10], CUADS [11], and DREAMER [12]. These datasets are combined into a single normalized dataset comprising 3,216 trials from 119 participants. Training across samples from multiple datasets requires each model to generalize across variations in ECG recording equipment, environments, and participant demographics during training. To our knowledge, this study is the first to utilize inter-dataset training and validation to emphasize generalization, as opposed to multi-dataset cross-validation, which focuses on adaptability.

To facilitate this benchmark study, we introduce two open-source frameworks: Affective Research on Representations and Classifications (ARRC, "*ark*") [17] and Affective Research Dataset Toolkit (ARDT, "*art*") [16].

This study makes the following key contributions:

1) Comparative Analysis of Deep Learning Models in AER: A controlled evaluation of prominent AER architectures, using ARRC and ARDT to perform a standardized benchmark that prioritizes model generalization over cross-dataset adaptability.
2) Affective Research on Representations and Classifications (ARRC, *"ark"*): A benchmark framework for fair, reproducible model comparisons in AER. ARRC also supports new model development, architectural experimentation, and hyperparameter optimization, making it a versatile tool for AER research.

3) Affective Research Dataset Toolkit (ARDT, *"art"*): A framework for working with AER datasets that enables inter-dataset training and validation. ARDT also provides a consistent interface for interacting with individual AER datasets, standardizing access regardless of differences in schema or data format.

This work builds on and extends a prior doctoral dissertation [26], which developed the foundational ARRC and ARDT frameworks and explored real-time, privacy-centric affect monitoring on embedded wearable devices. While the dissertation focused on the end-to-end system design—including on-device inference, privacy considerations, and embedded deployment—this paper isolates and expands the model benchmarking component into a standalone, reproducible comparative study. Specifically, we extend the original work by (1) providing a more detailed and controlled evaluation of five deep learning architectures under standardized conditions, (2) incorporating 10-fold cross-validation with hyperparameter optimization to strengthen the statistical rigor of the comparison, and (3) presenting the ARRC and ARDT tools as independent, open-source contributions for the broader AER research community.

## II. Prior Work

Prior work has shown that CNN-based deep learning models significantly improve prediction accuracy for AER problems. Various studies have proposed different CNN architectures. Table I summarizes the classification metrics of three high-performing models in recent literature.

Each of these models consists of feature extraction and final output components, with the latter comprising one or more fully connected layers that yield either a regression or classification result. In this study, we focus on the feature extraction architecture, and use a common output component to fairly evaluate the differences between them.

TABLE I
AER Model Comparison

| | Training Dataset | Evaluation Dataset | Arousal Accuracy | Valence Accuracy |
|---|---|---|---|---|
| **CNN+LSTM** | AMIGOS | AMIGOS | - | 79.0% |
| | DREAMER | DREAMER | - | 70.0% |
| **PETSFCNN** | DREAMER | DREAMER | 97.56% | 96.3% |
| **T-CNN** | ASCERTAIN | ASCERTAIN | 94.83% | 93.10% |
| | ASCERTAIN + DREAMER | DREAMER | **98.65%** | **97.30%** |

### A. CNN+LSTM

Harper et al. proposed a feature fusion technique wherein features are extracted from the input data through two parallel streams, as shown in Figure 1. The first stream is a traditional CNN architecture, with four stacked convolutional layers that extract local patterns across the duration of the input time series data, followed by a pooling layer. Each convolution layer uses he-normal kernel initialization and relu activation and is followed by a 50% dropout. The second stream is a bi-directional long short-term memory (LSTM) with 32-units, which extracts past- and future-sequence structures from the input. The outputs of each stream are concatenated to form the extracted feature vector. In the original study, the authors used a final single-neuron fully connected layer to produce a regression value estimating valence. Additionally, they utilized a Bayesian framework to model uncertainly during inference to improve final classification metrics. In this study, we focus on the feature-extraction capabilities of the CNN+LSTM model architecture, omitting the final regression layer as well as the Bayesian uncertainty during the validation phase. The study authors report 79% classification accuracy on the valence axis.

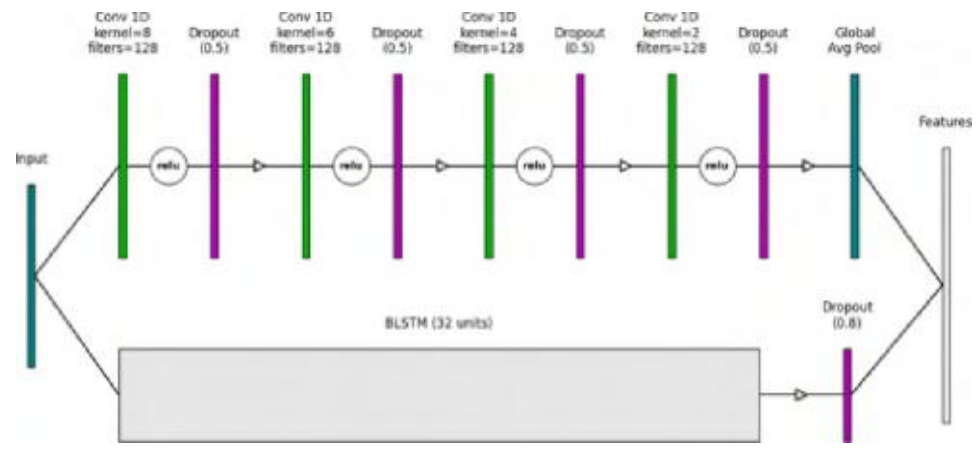

Fig. 1. **CNN-LSTM Feature Extraction.** ECG input signal is simultaneously processed through (a) four 1x1 convolutions, each followed by a 50% dropout, followed by global average pooling, and (b) a bi-direction LSTM followed by an 80% dropout. The outputs are concatenated to form the extracted feature vector.

In the original study, the authors manually extracted inter-beat interval (IBI) data from the ECG signals and used that as the time-series input to the CNN+LSTM. They note that the reduced amount of input data compared to ECG makes training more difficult. We infer from this that the model is applicable to the ECG signal data as well and proceed as such in our benchmark study.

### B. Temporal CNN

Finally, we include a Temporal CNN. The Temporal CNN model is designed as a CNN-based autoencoder, with both the number of filters and the down-sampling rate increasing with depth. Each layer is implemented as a residual block featuring exponentially increasing dilated causal convolutions. The use of stacked, dilated, causal convolutions enable the T-CNN model to extract both local and global temporal features of the input signal, achieving performance comparable to the feature-fusion approaches discussed above, without added complexity. The residual blocks employ a gated activation function popularized by the Wavenet architecture [13] for feature extraction from time-series audio data. The architecture is summarized in Figure 2.

This is a simplified version of the model from the original study, where each residual block included a multi-layer temporal CNN to generate its residual output. In the current approach, each residual layer consists of only a single layer of the overall temporal CNN structure.

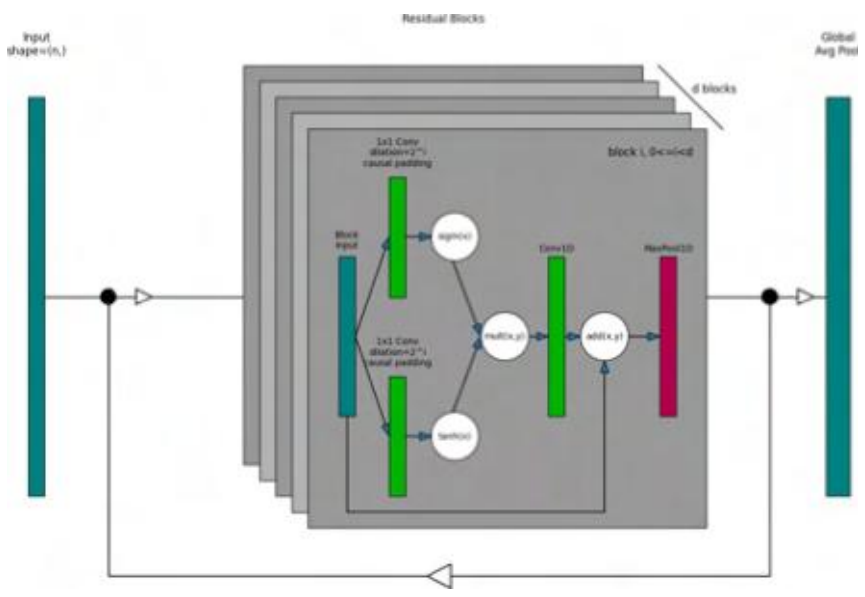

Fig. 2. **Temporal CNN Resnet Architecture.** ECG input signal passes through $d$ residual blocks, where each block performs a gated activation using 1x1 convolutions with dilation=$2^i$ where $0 <= i < d$, followed by a 1x1 convolution, addition of the residual, and max pooling.

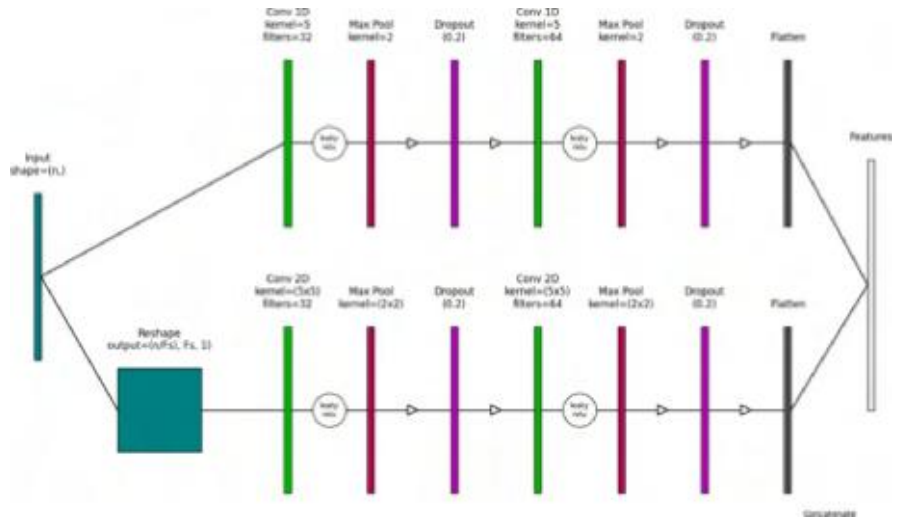

Fig. 3. **PETSFCNN Feature Extraction.** ECG input signal is processed through a 1D convolution network and simultaneously reshaped into a 2D structure and processed through 2D convolution layers. The outputs are concatenated to form the extracted feature vector.

## C. PETSFCNN

Parallel Extraction of Temporal and Spatial Features using CNN (PETSFCNN) is a CNN-based feature fusion model for ECG-based emotion classification. In PETSFCNN, the input ECG signal is processed through both 1D and 2D CNN networks. The 1D CNN extracts local temporal features, while the 2D CNN extracts spatial features of the ECG signal. To prepare the input for the 2D CNN, the authors segment the ECG into 1-second non-overlapping windows, stack them, and rescale the values between $[0, 1.0]$, interpreting them as a greyscale image.

Both the 1D and 2D CNN streams are comprised of two convolution blocks, with each block containing a convolutional layer with leaky-relu activation, max pooling with kernel size 2, and a 20% dropout. The convolution layer in the first block of each stream uses 32 filters and a kernel size of 5, and the convolution layer in the second uses 64 filters and a kernel size of 3. The outputs of each are concatenated together to form the final extracted feature vector. The architecture is shown in Figure 3.

PETSFCNN applies min-max scaling to the input ECG signal, which we retain in our implementation for this study, though it is not explicitly shown.

This enables the model to extract global structural patterns from the time-series data through 2-D convolutional layers, in addition to the local features obtained from the 1-D convolutional branch.

## D. Baseline Models

In order to provide a fair baseline comparison, we include two additional CNN-based models in this study. The first is the traditional 1D CNN architecture comprised of layered blocks containing convolution and pooling operations. We replicate the 1D-CNN branch of the PETSFCNN model for this purpose. The second is a ResNet architecture comprised of residual blocks. Each residual block performs convolution and pooling operations before adding the block's input to the block's output to feed forward to the next block in sequence. This is essentially the T-CNN architecture, except each block utilizes ReLU instead of gated activation, and the convolutions use the *same* padding rather than *causal* without kernel dilation.

## E. Limitations of Prior Work

Studies reporting results for various deep learning architectures have primarily focused on classification metrics to evaluate performance, directly comparing their findings with those from previous work. This approach creates an illusion of incremental improvements while overlooking the influence of ECG signal quality across datasets, signal pre-processing methods, and variations introduced by different deep learning frameworks. The small size of individual datasets also complicates direct comparisons between studies. For example, using a 10% leave N-participant-out validation split on the DREAMER dataset would yield a validation set with $\lfloor 23 * 0.1 \rfloor * 18 = 36$ samples. With 36 samples in the validation set, 1 sample accounts for $1/36 = 0.027$ points, or $\pm 2.7\%$, of the total accuracy metric. While methods like $k$-fold cross-validation mitigate this issue, attributing a single misclassification to random chance in the validation splits or variance in signal pre-processing is reasonable. Consequently, the 1% difference in classification accuracy between T-CNN and PETSFCNN, for example, falls within an expected margin of error.

Furthermore, each of the AER models discussed has been trained and validated against a single dataset. Cross-validation with a second dataset shows the model's ability to adapt to variations introduced by recording equipment and conditions but does not directly support the model's ability to generalize across them.

# III. Data Selection and Processing with ARDT

One primary goal of this study is to include samples from various publicly available AER datasets. Individual datasets use different schemas and formats, making data preparation and processing for model training time-consuming. To address this, we developed the Affective Research Dataset Toolkit (ARDT). ARDT is a Python API that offers a common interface to interact with the underlying datasets, abstracting implementation details and differences.

A detailed overview of the ARDT library is beyond the scope of this benchmark study. However, the source code is available in a public GitHub repository [16].

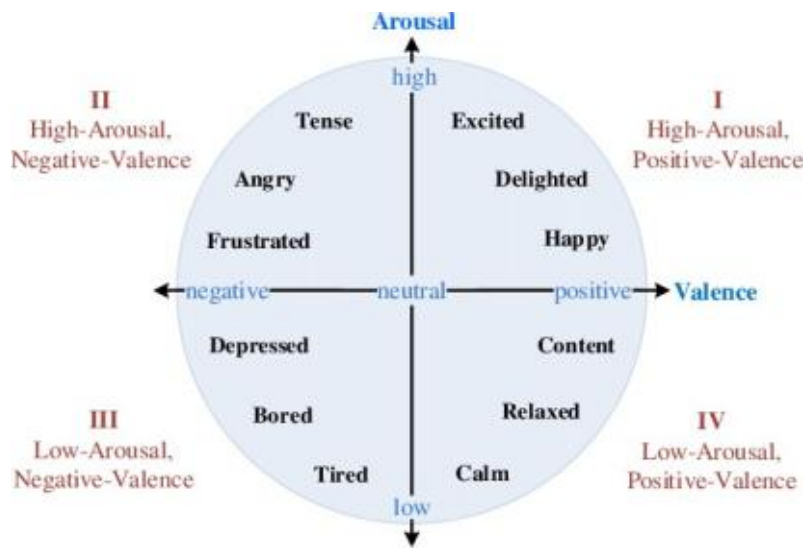

Fig. 4. **Arousal-Valence plane** for classifying emotion on two axes. Arousal as a measure of excitation being 'high' or 'low', and valence a measure of pleasantness, being 'positive' or 'negative'.

In this benchmark, we use samples from the ASCERTAIN, CUADS, and DREAMER datasets. ASCERTAIN and DREAMER were selected due to their use in prior research. They are supplemented by CUADS, a publicly available dataset compiled during this study's design. Each dataset measures emotion on the Arousal-Valence plane, as shown in Figure 4.

ASCERTAIN, CUADS, and DREAMER each use a dimensional model of human emotion known as the Arousal/Valence plane [14]. In this model, emotions are classified along two axes: Arousal, which measures excitation as either ”high” or ”low,” and Valence, which measures pleasantness as either ”positive” or ”negative.”

Each of the datasets have significant class imbalances in their participant responses, with 84.1% of ASCERTAIN, 59.6% of CUADS, and 72.3% of DREAMER representing high arousal emotional states in Quadrants I and II. The classifications are summarized in Table II. Large class imbalances can introduce prediction bias due to the disproportionate influence on model training [15]. ARDT provides methods to over- or under-sample ARDT datasets to provide perfectly balanced data for training and validation.

Finally, ARDT provides a simple API to generate training and validation splits with N participants excluded. For this benchmark study, we create 80% training and 20% validation splits, followed by undersampling.

The final step in preparing the data for use is to apply a signal-filtering pipeline. ARDT provides an API for implementing signal processors that are easily chained together to form a complete, reusable pipeline that is applied to the signal data in each dataset trial. For this benchmark, we construct an ARDT signal processor chain that performs the following actions in order:

1) Model baseline wander by applying 600ms and 200ms median filters, sequentially. [21]
2) Subtract baseline wander from the ECG signal.
3) Eliminate AC powerline noise, EMG interference, and other high frequency noise by apply a $12^{th}$ order low-pass butterworth filter with 35 Hz cut-off.
4) Downsample the ECG signal from 256 Hz to 128 Hz.
5) Apply z-score normalization.

TABLE II
SUMMARY OF AER DATASET SIZE, BY A/V QUADRANT. SIGNIFICANT CLASS IMBALANCE PRESENT IN HIGH-AROUSAL CLASSES.

| Dataset | Quadrant I | II | III | IV | Total |
|---|---|---|---|---|---|
| **ASCERTAIN** | 1,038 | 719 | 88 | 243 | 2,088 |
| **CUADS** | 243 | 183 | 150 | 138 | 714 |
| **DREAMER** | 176 | 124 | 37 | 77 | 414 |
| **Total** | 1,457 | 1,026 | 275 | 458 | 3,216 |
| **Percent** | 45.3% | 31.9% | 8.6% | 14.2% | 100% |

6) Pad or trim the signal to 45 seconds. For signals longer than 45 seconds, the signal is truncated to the final 45 seconds of data. For signals shorter than 45 seconds, the signal is left-padded with 0 values.

The implementation of the pipeline is out of scope for this paper, but is available in the study's source code [25]. For multi-channel ECG signals, the pipeline is applied to each channel individually.

## IV. BENCHMARK DESIGN WITH ARRC

The Affective Research on Representations and Classifications (ARRC) is an open-source framework for AER model development. It is written using the multi-backend Keras 3.0 API [18] with support for both Tensorflow [19] and PyTorch [20] backends. The core of ARRC is the `ARRCModel` class that encapsulates a user-defined feature extractor. `ARRCModel` provides an optional classification head, enabling it to be used for metric learning and classification problems alike. Loss functions can be applied to embedding output or classification output layers separately or as weighted losses applied to both simultaneously. The ARRC source code [17] includes implementations of the feature extractors described in Section II, and is compatible with any Keras, Tensorflow or PyTorch loss function.

ARRC also provides several custom layers used for data augmentation during training. The available data augmentations include additive Gaussian noise, random time shift, and random amplitude scaling.

ARRC is used to ensure that each model architecture is assessed fairly by providing a common base model with a consistent classification head, and consistent application of weighted losses and metrics.

## V. TRAINING METHODOLOGY

### *A. Benchmark Setup*

For this benchmark study, each feature extraction model is reproduced as described in Section II. Models are trained to predict Arousal only, using a 1-class binary classification with sigmoid activation. An output value of 1 indicates high-arousal emotions in Quadrants I and II, while 0 indicates low-arousal emotions in Quadrants III and IV, as shown in Figure 4.

We apply two weighted loss functions: a contrastive loss with online pair mining for the embeddings and a binary cross-entropy loss for classification. The contrastive loss enhances class separation in the embedding space [24], while the cross-entropy loss promotes positive clustering.

Each model is trained using the Adaptive Moment Estimation (Adam) optimizer with an initial learning rate $0.0001$. A cosine-decay learning rate scheduler is utilized with a period of 250 epochs. Models are trained for up to 1,000 epochs, with early stopping if the validation accuracy does not improve within 250 epochs. Upon completion of training, weights are restored from the epoch with the highest validation accuracy, and classification metrics are collected.

10-fold cross-validation is employed. Classification metrics for each model are presented as the mean results of the 10 iterations. In each fold, the training and validation splits are under-sampled to tackle the significant class imbalance in the datasets, as shown in Table II. To ensure consistency, the dataset splits are generated in advance and reused for each model. Due to random participant selection during split generation, each split has a varying number of trials. The mean number of trials in the training splits is 558.4 ($\sigma = 35.2$), with 139.6 per quadrant ($\sigma = 8.8$). The mean number of trials in the validation splits is 153.6 ($\sigma = 35.2$), with 38.4 per quadrant ($\sigma = 8.8$).

### B. Regularization and Data Augmentation

Training deep learning models for AER problems is particularly challenging. The subjective nature of participants' self-reports on emotional experiences results in classes with poorly defined separation boundaries. Additionally, the small size and class imbalances of the datasets lead to overfitting.

To eliminate variance from participant responses, we conduct all training and validation using the pre-validated target labels from each dataset.

To address overfitting, batch normalization and dropout regularization layers for each model, as described in their original papers. In addition to these regularization techniques, we use several standard data augmentations techniques for time-series input data [22], [23]. Random time shifting left-pads the ECG signal to shift it forward in time by up to 10% of its total duration. Random amplitude scaling scales the input ECG signal $\pm 10\%$. Finally, we add small amounts of Gaussian noise to the inputs signals. These augmentations are implemented in ARRC as custom layers that are applied to the model's input. Each augmentation randomly selects a subset of signals from the input batch and applies its transformation to them.

## VI. Results

Since we did not take additional steps to tune model performance for a specific application, the results presented should not be interpreted as suggesting that any one model performs better than another. Each model is discussed in relation to its baseline for comparison. The classification metrics collected are averaged over the 10-fold cross-validation and summarized in Table III, with ROC curves shown in Figure 5.

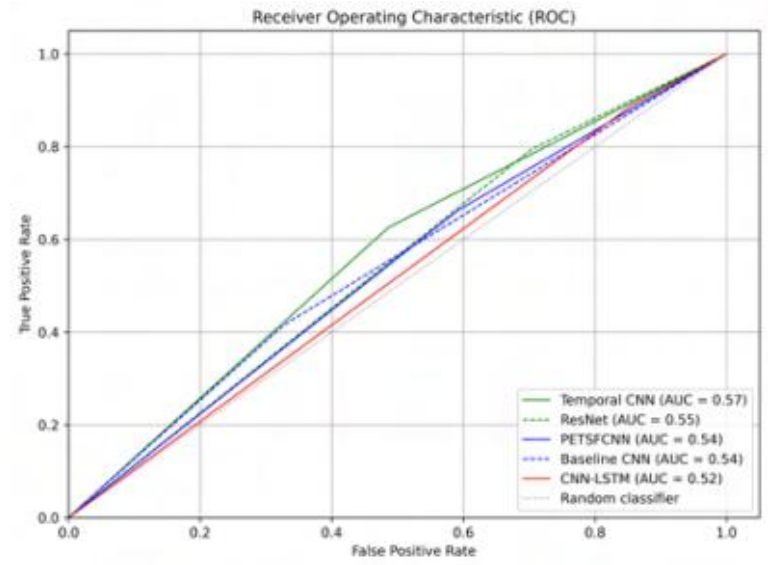

Fig. 5. **ROC-AUC Curves** for models in this benchmark.

### A. Baseline Models

The baseline CNN and ResNet architectures performed similarly, achieving mean classification accuracies of 54.4% and 54.5% respectively.

The baseline CNN architecture has $11M$ trainable parameters, with a memory footprint of $45MB$. Model inference performs $24M$ floating-point operations (FLOPs), per input sample. In each of our tests, an input sample is 45 seconds sampled at 128Hz. The ResNet model has $68K$ trainable parameters requiring $268KB$ memory, and performs $23.4M$ FLOPs per input sample.

The baseline CNN is the 1-D branch of the PETFSCNN architecture. It's large capacity, relative to ResNet, is due to the handling of the output after the final convolutional layer. PETSFCNN flattens the final output, where ResNet applies global average pooling which greatly reduces the size of the final feature vector.

### B. CNN-LSTM

The CNN-LSTM model is the sole outlier with no direct baseline for comparison. Conceptually, the CNN-LSTM model is similar to the T-CNN in that it seeks to extract both local and global temporal features, and we anticipated similar performance between the two. However, the CNN-LSTM only achieved 51.8% classification accuracy with an AUC of $0.52$.

In the original study, the authors used interbeat-interval (IBI) data extracted from ECG signals as input to the model. They noted the tradeoff between the loss of data carried by the ECG signal and the elapsed-time performance gained by training on IBI intervals instead. We inferred from this that their model was suitable for use with ECG input signals, but the results suggest that significant fine-tuning is required. At a minimum, the size of the LSTM layer should be adjusted to account for the difference in sampling rate between the ECG and IBI signals. Although not reported in this study, we were able to successfully reproduce the results of the original study using IBI extracted from the ECG using ARDT.

The CNN-LSTM model has $219K$ trainable parameters using $856KB$ memory. Despite its small size, the LSTM has high computational complexity, requiring $98M$ FLOPs per input sample.

TABLE III
SUMMARY OF PERFORMANCE METRICS FOR EACH MODEL, WITH MODEL SIZE AS THE TOTAL NUMBER OF TRAINABLE PARAMETERS, THE MEMORY FOOTPRINT NEEDED FOR INFERENCE, AND MODEL COMPLEXITY AS THE TOTAL NUMBER OF FLOPS PER INPUT SAMPLE.

| Metric | 1-D CNN | ResNet | CNN-LSTM | T-CNN | PETSFCNN |
|---|---|---|---|---|---|
| Acc. | 0.544 | 0.545 | 0.518 | 0.570 | 0.535 |
| F1 | 0.559 | 0.530 | 0.511 | 0.563 | 0.528 |
| AUC | 0.540 | 0.550 | 0.520 | 0.570 | 0.540 |
| # Params | $11M$ | $68K$ | $219K$ | $178K$ | $14M$ |
| Memory | $45MB$ | $268KB$ | $856KB$ | $698KB$ | $45MB$ |
| FLOPs | $24M$ | $23.4M$ | $98M$ | $20M$ | $76M$ |

### C. Temporal CNN

The Temporal CNN model performed significantly better than its ResNet baseline, achieving a mean accuracy of $57.0\%$, with an AUC of $0.57$. The T-CNN model is a residual block architecture, but utilizes gated activations with skip connections, and dilated temporal convolutions to capture global temporal patterns and dependencies.

This model has $178K$ trainable parameters, requires $698KB$ RAM, and performs $20M$ FLOPs per input sample during forward pass.

### D. PETSFCNN

It is interesting to observe that in this benchmark, PETSFCNN achieved a slightly lower classification accuracy, $0.535$, than its 1-D CNN baseline, with both having an ROC-AUC score $0.540$ suggesting identical classification performance. The difference in accuracy score is likely due to the increased capacity of PETSFCNN over its baseline because of the 2D-CNN branch, which would delay convergence. This is supported by the identical ROC-AUC scores, which suggest the 2-D branch adds no predictive value to the model architecture.

We offer two possible explanations for this. First, it is likely that the 2-D representation of the ECG needs further processing before being input into the CNN layers. After reshaping, we apply a min-max rescaling to ensure its values are in the range $[0, 1]$. No additional processing was noted in the original paper. Second, we note that the 2-D representation contains the same time-domain information as the original 1-D signal input stream. It is likely that an alternative representation, such as a short-time Fourier transform (STFT) spectrogram, which plots the time domain against the frequency domain, may have yielded better results. This is left as an exercise for future work.

PETSFCNN has a $14M$ trainable parameters with a $45MB$ memory footprint, and performs $76M$ FLOPs per input sample. The large memory footprint is due to the flattening and concatenation of the outputs from the 1-D and 2-D convolutional branches. The use of a global pooling layer instead would make it comparable the other models. The 2-D CNN branch, however, adds significant computational complexity over its 1-D baseline.

## VII. DISCUSSION

One general observation is that larger and more complex architectures, such as the CNN-LSTM and PETSFCNN, may offer higher representational capacity but also carry significant memory and computational costs. These models may be more appropriate for high-performance computing environments or cloud-based systems where inference latency and energy constraints are less critical. In contrast, the Temporal CNN (T-CNN), while delivering comparable performance in terms of accuracy and AUC, achieves this with fewer parameters and lower FLOPs per input sample. This suggests it may be better suited for embedded or resource-constrained applications, such as real-time affective monitoring on wearable devices.

TABLE IV
PERFORMANCE METRICS FOR T-CNN AND PETSFCNN WHEN TRAINED AND VALIDATED ONLY ON THE CUADS DATASET. COMPARE TO INTER-DATASET RESULTS IN TABLE III.

| Metric | T-CNN | PETSFCNN |
|---|---|---|
| Accuracy | 0.598 | 0.559 |
| F1-score | 0.596 | 0.569 |
| AUC | 0.600 | 0.560 |

To illustrate the limitations of intra-dataset training, we conducted a small-scale experiment in which two representative models (T-CNN and PETSFCNN) were trained and validated exclusively on the CUADS dataset. As summarized in the Table IV, both models achieved higher accuracy and AUC under these conditions compared to their performance in the inter-dataset setting. This result supports our central claim that intra-dataset validation may overestimate generalization performance due to reduced variability in recording conditions and subject pool.

Several limitations of this study should be acknowledged. First, while the models were evaluated under standardized conditions, we did not perform extensive hyperparameter tuning for each architecture beyond initial grid searches. Architecture-specific fine-tuning would change the relative rankings of the models. It is important to highlight that the Temporal CNN model is the basis of prior work [5], which likely introduced unintended performance bias through default hyperparameter selections used during data collection.

Second, the decision to evaluate each model on arousal classification only—rather than on both arousal and valence—simplifies the emotional classification task and may not fully reflect real-world use cases where emotional states span a two-dimensional plane. Further, we perform training and validation using each dataset's pre-validated target labels rather than the participant's responses. Previous work using the CUADS dataset has shown moderate inter-rater agreement on video labels (e.g., Randolph's Kappa), supporting the use as target labels in the present study to eliminate the need to measure variance and confidence among the participant ratings. However, the predetermined class labels are not representative of the participant's experience during the recording session, which is recorded by the physiological sensors, and thus likely contribute to the low overall classification accuracies achieved.

Finally, while the use of ARRC and ARDT improves transparency and reproducibility, all implementations were reproduced based on descriptions from the original papers. Any discrepancies in architectural details not fully specified or misinterpreted from those sources may influence outcomes.

## VIII. Conclusion

This study presented a standardized benchmarking framework for ECG-based emotion recognition, enabling direct comparison of deep learning architectures under identical training, validation, and data processing conditions. By introducing two open-source tools—ARRC and ARDT—we provide the research community with extensible, reproducible frameworks for model development and inter-dataset evaluation in automated emotion recognition.

Our results demonstrate that architectural complexity alone does not guarantee improved classification performance, as evidenced by the Temporal CNN achieving competitive accuracy with significantly fewer parameters and lower computational cost than larger models such as CNN-LSTM and PETSFCNN. The use of inter-dataset training and validation, combining ASCERTAIN, DREAMER, and CUADS into a single corpus, provided a more stringent test of generalization than conventional within-dataset evaluation, as supported by the higher intra-dataset accuracies observed in our comparative experiment.

## References

[1] N. Fragopanagos and J. G. Taylor, "Emotion recognition in human-computer interaction," *Neural Networks*, vol. 18, no. 4, pp. 389–405, May 2005.

[2] D. Sander, D. Grandjean, and K. R. Scherer, "A systems approach to appraisal mechanisms in emotion," *Neural Networks*, vol. 18, no. 4, pp. 317–352, 2005.

[3] D. S. Hammad and H. Monkaresi, "ECG-based emotion detection via parallel-extraction of temporal and spatial features using convolutional neural network," *Traitement du Signal*, vol. 39, no. 1, pp. 43–57, Feb. 2022.

[4] R. Harper and J. Southern, "A Bayesian deep learning framework for end-to-end prediction of emotion from heartbeat," *IEEE Trans. Affect. Comput.*, vol. 13, no. 2, pp. 985–991, 2020.

[5] T. C. Sweeney-Fanelli and M. Imtiaz, "ECG-based automated emotion recognition using temporal convolution neural networks," *TechRxiv*, 2024.

[6] H. Zhu, M. Akrout, B. Zheng, A. Pelegris, A. Jayarajan, A. Phanishayee, B. Schroeder, and G. Pekhimenko, "Benchmarking and analyzing deep neural network training," in *Proc. IEEE Int. Symp. Workload Characterization (IISWC)*, 2018, pp. 88–100.

[7] B. Pyakillya, N. Kazachenko, and N. Mikhailovsky, "Deep learning for ECG classification," in *J. Phys.: Conf. Ser.*, vol. 913, no. 1, Oct. 2017.

[8] S. S. Alahmari, D. B. Goldgof, P. R. Mouton, and L. O. Hall, "Challenges for the repeatability of deep learning models," *IEEE Access*, vol. 8, pp. 211860–211868, 2020.

[9] Y. S. Can and C. Ersoy, "Smart affect monitoring with wearables in the wild: An unobtrusive mood-aware emotion recognition system," *IEEE Trans. Affect. Comput.*, vol. 14, no. 4, pp. 2851–2863, 2023.

[10] R. Subramanian, J. Wache, M. K. Abadi, R. L. Vieriu, S. Winkler, and N. Sebe, "ASCERTAIN: Emotion and personality recognition using commercial sensors," *IEEE Trans. Affect. Comput.*, vol. 9, no. 2, pp. 147–160, Apr. 2018.

[11] T. C. Sweeney-Fanelli, A. Ahmed, and M. H. Imtiaz, "Descriptor: Clarkson University Affective Research Dataset (CUADS)," *IEEE Data Descriptions*, pp. 1–10, 2025.

[12] S. Katsigiannis and N. Ramzan, "DREAMER: A database for emotion recognition through EEG and ECG signals from wireless low-cost off-the-shelf devices," *IEEE J. Biomed. Health Inform.*, vol. 22, no. 1, pp. 98–107, Jan. 2018.

[13] A. van den Oord, S. Dieleman, H. Zen, K. Simonyan, O. Vinyals, A. Graves, N. Kalchbrenner, A. Senior, and K. Kavukcuoglu, "WaveNet: A generative model for raw audio," *arXiv preprint arXiv:1609.03499*, Sep. 2016.

[14] J. A. Russell, "A circumplex model of affect," *J. Personality Social Psychol.*, vol. 39, no. 6, pp. 1161–1178, 1980.

[15] T. Meng, Y. Shou, W. Ai, N. Yin, and K. Li, "Deep imbalanced learning for multimodal emotion recognition in conversations," *IEEE Trans. Artif. Intell.*, vol. 5, no. 12, pp. 6472–6487, 2024.

[16] T. C. Sweeney-Fanelli, "Affective Research Dataset Toolkit (ARDT) Python API for inter-dataset collaborative research," 2025. [Online]. Available: URL withheld for double-blind review

[17] Organization withheld for double-blind review, "Affective Research on Representations and Classifications (ARRC)," 2025. [Online]. Available: URL withheld for double-blind review

[18] F. Chollet *et al.*, "Keras," 2015. [Online]. Available: https://keras.io

[19] M. Abadi, A. Agarwal, P. Barham, E. Brevdo, Z. Chen, C. Citro, G. S. Corrado, A. Davis, J. Dean, M. Devin, *et al.*, "TensorFlow: Large-scale machine learning on heterogeneous systems," 2015. [Online]. Available: https://www.tensorflow.org/

[20] A. Paszke, S. Gross, F. Massa, A. Lerer, J. Bradbury, G. Chanan, T. Killeen, Z. Lin, N. Tomkins, A. Desmaison, *et al.*, "PyTorch: An imperative style, high-performance deep learning library," in *Advances in Neural Information Processing Systems*, 2019, pp. 8024–8035.

[21] Y.-L. Hsu, J.-S. Wang, W.-C. Chiang, and C.-H. Hung, "Automatic ECG-based emotion recognition in music listening," *IEEE Trans. Affect. Comput.*, vol. 11, no. 1, pp. 85–99, Jan. 2020.

[22] K. M. Rashid and J. Louis, "Times-series data augmentation and deep learning for construction equipment activity recognition," *Adv. Eng. Inform.*, vol. 42, 2019.

[23] Q. Wen, L. Sun, F. Yang, X. Song, J. Gao, X. Wang, and H. Xu, "Time series data augmentation for deep learning: A survey," in *Proc. Int. Joint Conf. Artif. Intell. (IJCAI)*, 2021.

[24] R. Hadsell, S. Chopra, and Y. LeCun, "Dimensionality reduction by learning an invariant mapping," in *Proc. IEEE Comput. Soc. Conf. Comput. Vis. Pattern Recognit.*, vol. 2, 2006.

[25] T. C. Sweeney-Fanelli, "ARRC/ARDT benchmark study source code repository," 2025. [Online]. Available: URL withheld for double-blind review

[26] T. Sweeney-Fanelli, "Real-time, privacy-centric embedded affect monitoring with wearable sensors," Order No. 31934646, Clarkson University, United States – New York, 2025.